# Synthetic extinction maps around intermediate-mass black holes in Galactic globular clusters

C. Pepe[1,2]⋆ and L. J. Pellizza[2,3]
[1] *Instituto de Astronomía y Física del Espacio, Av. Int. Guiraldes s/n, Ciudad Autónoma de Buenos Aires, Argentina*
[2] *Consejo Nacional de Investigaciones Científicas y Técnicas, Av. Rivadavia 1917, Ciudad Autónoma de Buenos Aires, Argentina*
[3] *Instituto Argentino de Radioastronomía, Camino Gral. Belgrano Km 40, Villa Elisa, Buenos Aires, Argentina*

## ABSTRACT

During the last decades, much effort has been devoted to explain the discrepancy between the amount of intracluster medium (ICM) estimated from stellar evolution theories and that emerging from observations in globular clusters (GCs). One possible scenario is the accretion of this medium by an intermediate-mass black hole (IMBH) at the centre of the cluster. In this work, we aim at modelling the cluster colour-excess profile as a tracer of the ICM density, both with and without an IMBH. Comparing the profiles with observations allows us to test the existence of IMBHs and their possible role in the cleansing of the ICM. We derive the intracluster density profiles from hydrodynamical models of accretion onto a central IMBH in a GC and we determine the corresponding dust density. This model is applied to a list of 25 Galactic GCs. We find that central IMBHs decrease the ICM by several orders of magnitude. In a subset of 9 clusters, the absence of the black hole combined with a low intracluster medium temperature would be at odds with present gas mass content estimations. As a result, we conclude that IMBHs are an effective cleansing mechanism of the ICM of GCs. We construct synthetic extinction maps for M 62 and $\omega$ Cen, two clusters in the small subset of 9 with observed 2D extinction maps. We find that under reasonable assumptions regarding the model parameters, if the gas temperature in M 62 is close to 8000 K, an IMBH needs to be invoked. Further ICM observations regarding both the gas and dust in GCs could help to settle this issue.

## 1 INTRODUCTION

The fate of interstellar medium (ISM) in globular clusters (GCs) is still an unresolved issue. Although stellar evolution theories predict between 10–100 $M_\odot$ of ISM, observations have failed to detect such quantities and they only have succeeded at placing upper limits well below the predicted values. For instance, van Loon et al. (2006) estimated upper-limits for a small sample of GCs and detected 0.3 $M_\odot$ of atomic hydrogen in M 15 while Freire et al. (2001) estimated a similar amount (0.1 $M_\odot$) of ionized hydrogen in the particular case of 47 Tuc. Later, van Loon et al. (2009b) determined additional, more stringent upper limits for another sample of GCs. In addition, sub-millimetric and infrared observations have been performed in the search for dust, yielding upper limits of $\sim 10^{-5}$–$10^{-2}$ $M_\odot$ (Barmby et al. 2009, see Priestley 2011 for a rather comprehensive review on the subject). In the particular case of M 15, Evans et al. (2003) reported a first detection of 5 $10^{-5}$ $M_\odot$. This result was later refined by Boyer et al. (2006) who estimated $\sim 10^{-5}$ $M_\odot$ of dust in this cluster. Thus, a cleansing mechanism that removes the ISM of GCs needs to be invoked.

Models intending to explain the discrepancy between stellar evolution theories and observations have been developed since the middle 1970s. Scott & Rose (1975) and Faulkner & Freeman (1977) studied an intrinsic cleansing mechanism by solving the steady-state flow equations in one dimension. In both cases the authors assume that the most evolved stars contribute to the ISM by means of their winds. However, Scott & Rose (1975) assume that there is sufficient stellar ultraviolet radiation to maintain hydrogen fully ionized and that this is the unique energy input to the gas system while Faulkner & Freeman (1977) consider the corresponding injection of energy and the consequent collisional ionization by electrons along with the associated radiative cooling of the gas. Under the same physical assumptions, the work of Faulkner & Freeman (1977) was later refined by VandenBerg & Faulkner (1977), who investigated the time-dependent equations. Finally, McDonald & Zijlstra (2015) showed that UV radiation from white dwarfs can efficiently clear the intracluster medium Yet, extrinsic mechanisms arising from the GC environment have also been considered. Frank & Geisler (1976) studied an analytical model concerning the effect of the GC moving through the Galactic halo medium as a sweeping mechanism. More recently, Priestley et al. (2011) developed the first 3D hydrodynamical simulation of such scenario. Accretion flows have also become an appealing scenario. Krause et al. (2012) stated that the power required to expel the gas from the GC can be provided by a coherent onset of accretion onto the stellar remnants of supernovae. Similarly, Scott & Durisen (1978) and Moore & Bildsten (2010) investigated the input of energy by hydrogen-rich explosions on accreting white dwarfs. Further, Leigh et al. (2013) argue that accretion onto stellar-mass black holes is an effective mechanism for rapid gas depletion. Finally, we (Pepe & Pellizza 2013, hereafter Paper I) developed a numerical model for the steady-state isothermal flow resulting from the constant injection of mass by the red giants in the cluster in the presence of an accreting intermediate-mass black hole (IMBH). We showed that a significant fraction of the ISM can be removed from the GC either via accretion onto the black hole in the inner regions or winds in the external regions. It is clear that efforts have been devoted to explain the cited discrepancy between theory and observations. However, due to the lack of measurements, little has been done in testing the density profiles predicted by the models, as an evaluation of the main hypotheses of the model under consideration. One goal of this work is to develop a method to this aim.

⋆ Currently at Instituto Argentino de Radioastronomía. e-mail: carolina@iar.unlp.edu.ar

All of the cited works focused on the gaseous component of the ISM. However, the dusty component is coupled to the gas in the ISM with a density two–five orders of magnitude lower, according to the cluster metallicity. This dust introduces differential reddening in the light of GC stars. Some authors have developed new techniques to deredden the CMDs and obtain high-precision maps of the observed differential reddening of a vast sample of GCs. Alonso García et al. (2011) developed a method in which they use photometric studies of main sequence, subgiant branch and red giant branch stars but, unlike other authors (see Piotto et al. 1999, for example), the extinction is calculated not on a predetermined grid but on a star by star basis. McDonald et al. (2009) and van Loon et al. (2009) also constructed reddening maps on a star by star basis fow $\omega$ Cen, exclusively. As a result, these authors obtain differential extinction maps which could be used to detect the ISM and trace its density profile. This would provide us with a new observable to test different cleansing mechanisms, in particular that involving an IMBH, since the existence of these objects is still a topic of debate.

Different kinds of observations and theoretical works have suggested that black holes with masses in the range $10^2 - 10^4$ $M_\odot$ exist in the Universe (Fabbiano 2006; Feng & Soria 2011; Magorrian et al. 1998). According to them, globular clusters are the main candidates to host these IMBHs. For this reason, GCs have become the main target of observations that search for these elusive compact objects.

Up to date, two different approaches have been used to constrain the presence as well as the mass of IMBHs in Galactic globular clusters. The first one relies on the dynamical effects of the putative IMBH on the surrounding stars. From the combinations of different dynamical models with both surface brightness and velocity dispersion profiles, several authors have managed to place constraints on the hypothetical IMBH mass. However, there is no consensus about the masses measured. The reader is referred to the references in Feldmeier et al. (2013) for an extensive list of observations and results. Not only different dynamical models have led to very conflicting results, but also, given the spatial resolution of current telescopes, these methods do not allow to distinguish between one single object and a collection of smaller objects.

The second approach used to detect IMBHs in GCs is related to the study of the emission due to the accretion process. The most significant observation was reported by Nucita et al. (2008) who claim the detection of an X-ray source (using *Chandra* and *XMM-Newton*) located in the centre of NGC 6388 with spectral properties consistent with an accreting IMBH. However, *Chandra* was not able to detect any source in the core of M 15 (Ho, Terashima & Okajima 2003). Also, radio observations have been performed since Maccarone (2004) stated that IMBHs would be more easily detected in radio. Maccarone, Fender & Tzioumis (2005) reported the detection of an IMBH of $\lesssim$ 600 $M_\odot$ in M 15 (using *Very Large Array*, VLA) while no central sources have been detected in the core of $\omega$ Cen (from *Australia Telescope Compact Array*, ATCA, observations). The result for M 15 was later challenged by Bash et al. (2008) who failed to detect the core of this GC. Further, different upper limits on the IMBH mass have been reported by several authors for all of these GCs (e.g. Maccarone & Servillat 2010; Lu & Kong 2011; Strader et al. 2012; Cseh et al. 2010). The dependence of these results on the model used to describe how accretion onto the IMBH proceeds is discussed in Paper I.

It is clear that the state-of-the-art of this topic is complex and efforts should be made in order to resolve this issue. In this work, we present synthetic extinction maps of globular clusters from density profiles of models with and without IMBHs This allows us to investigate the reliability of the physical hypotheses underlying this particular cleansing mechanism. We choose the density profiles emerging from the model presented in Paper I, aiming at establishing the presence of an IMBH from observations of the gas content and extinction in globular clusters. In Sect. 2 we state the basic equations and describe the construction method in detail. Our sample consists of the 25 Galactic GCs where searches of IMBHs have been performed so far (see Table 1). In Sect. 3, we present our results for the density profiles, gas mass and extinction maps for different ISM parameters in our sample. Finally, in Sect. 4 we discuss the relevance of our results and the implementation of this method to other ISM models while in Sect. 5 we present our conclusions.

**Table 1.** List of globular clusters where IMBHs have been searched for, according to the literature, ordered by increasing $\sigma$. The maps of this work were constructed for this sample. The cluster parameters $r_0$ and $\rho_0$ were taken from Harris (1996) while $\sigma = \sqrt{4\pi G \rho_0 r_0^2}$ was calculated using the former parameters.

| ID | $r_0$ (pc) | $\rho_0$ ($M_\odot$ pc$^{-3}$) | $\sigma$(km s$^{-1}$) |
|---|---|---|---|
| NGC 6535 | 0.712 | 2.19E+2 | 0.82 |
| NGC 6397 | 0.03345 | 5.75E+5 | 1.99 |
| NGC 6652 | 0.2908 | 3.02E+4 | 3.98 |
| NGC 5694 | 0.6108 | 8.91E+3 | 4.54 |
| M 10 | 0.9855 | 3.47E+3 | 4.57 |
| NGC 5694 | 0.6186 | 8.91E+3 | 4.60 |
| NGC 6752 | 0.1978 | 1.10E+5 | 5.16 |
| M 79 | 0.6003 | 1.20E+4 | 5.18 |
| M 13 | 1.2804 | 3.55E+3 | 6.09 |
| M 5 | 0.9599 | 7.59E+3 | 6.58 |
| NGC 6402 | 2.1371 | 2.29E+3 | 8.05 |
| M 28 | 0.3839 | 7.24E+4 | 8.14 |
| NGC 5286 | 0.9594 | 1.26E+4 | 8.48 |
| M 80 | 0.4363 | 6.18E+4 | 8.54 |
| NGC 1851 | 0.3167 | 1.23E+5 | 8.75 |
| NGC 5824 | 0.5602 | 4.07E+4 | 8.90 |
| 47 Tuc | 0.4712 | 7.59E+4 | 10.22 |
| $\omega$ Cen | 3.5849 | 1.41E+3 | 10.61 |
| M15 | 0.4235 | 1.12E+5 | 11.17 |
| NGC 6440 | 0.3461 | 1.74E+5 | 11.36 |
| NGC 2808 | 0.6981 | 4.57E+4 | 11.76 |
| M 54 | 0.6937 | 4.90E+4 | 12.09 |
| M 62 | 0.4351 | 1.45E+5 | 13.03 |
| NGC 6388 | 0.3455 | 2.34E+5 | 13.18 |
| NGC 6441 | 0.4386 | 1.82E+5 | 14.74 |

## 2 THE METHOD

### 2.1 Reddening: Basic equations

In this Section we describe the basic equations underlying the method. It has been designed so as to calculate the reddening value for every single star in the cluster. The details concerning the binning and building of the map are explained in the forthcoming section.

The reddening maps constructed in this work show the distribution of the colour excess

$$E_{B-V} = (B - B_0) - (V - V_0) = A_B - A_V. \qquad (1)$$

where $A_B$ and $A_V$ are the absorption values for the B and V bands, respectively and were chosen in order to obtain maps comparable to those of Alonso García et al. (2011). Following Cardelli, Clayton & Mathis (1989), we assume a standard reddening law with $R_V = 3.1$, according to observational fits.

Here, we study the emission of each single star in the cluster and assume that there are no sources of radiation other than the star itself. Then, from the classic transport equation

$$\frac{dI_\nu}{ds} = -\kappa_\nu I_\nu, \qquad (2)$$

where $I_\nu$ is the radiation intensity and $\kappa_\nu$ the extinction coefficient, $s$ the distance travelled by the stellar light within the GC and the emission coefficient was already set to zero. Assuming pure absorption, we obtain the absorption law

$$I_\nu(s^*) = I_{\nu,0} \exp\left( - \int_{s_{min}}^{s*} \kappa_\nu ds \right). \qquad (3)$$

where $s*$ is the position of the star and $s_{min}$ the edge of the cluster closest to the observer, both along the line of sight. The extinction coefficient can be modeled as

$$\kappa_\nu = n\sigma_e^\nu, \qquad (4)$$

where $n = \rho/m_p$ is the particle density of the dust distribution and $\sigma_e^\nu$ is the cross section of the extinction process. Only absorption effects are considered whereas scattering processes are neglected.

Since the radiation intensity and the observed magnitude $m$ are related through $m = -2.5 \ \log_{10} I_\nu + C_\nu$, the absorption $A_B$ can be written as

$$A_B = 1.086 \ \sigma_e^B \int_{s_{min}}^{s*} n(s) ds. \qquad (5)$$

Therefore, Eq. 1 can be written in terms of the integral of the particle density profiles

$$E_{B-V} = 1.086 \ \sigma_e^V \left( \frac{A_B}{A_V} - 1 \right) \int_{s_{min}}^{s*} n(s) ds, \qquad (6)$$

where Eq. 5 has been used to replace the ratio $\sigma_e^B/\sigma_e^V$. The coefficient $A_B/A_V$ is taken from Cardelli, Clayton & Mathis (1989), who found a correlation between $\langle A(\lambda)/A(V)\rangle$ and $R_V$ (the reader is referred to that article for the complete expressions). The cross section $\sigma_e^V = q\pi a^2$ is assumed to be a geometrical parameter, where $a$, lying in the range 1Å$-1$ $\mu$m (Weingartner & Draine 2001) is the radius of a dust particle and the factor $q \sim 0.1$ determines the efficiency of the absorption process. Combining the value of the particle radius $a$ with the mean density of dust particles ($\rho_p \sim 3$ g cm$^{-3}$, Draine 2004), the particle mass can be calculated. A remarkable fact is that, given the dependence of $\sigma_e^\nu$ and $m_p$ on the radius of the particle $a$, the colour excess $E_{B-V}$ depends only on the ratio $q/a$.

Thus, given the position $s^*$ of a star in the cluster we can obtain an excess colour estimate $E_{B-V}$ by integrating the dust density profile up to $s^*$ as stated in Eq. 6. The construction details of the maps are stated below.

### 2.2 Construction of the maps

We used a Monte Carlo (MC) simulation to generate the synthetic stellar distribution of a given globular cluster. A Miocchi (2007) model was assumed for the potential of the cluster. The integration of such model leads to the radial stellar distribution used in the MC model. Once the tridimensional position of every single star is known, we project it on the plane of the sky and characterize its position with its transversal distance to the cluster centre $R_T$ and the depth $s^*$, which is the distance to the plane of the sky that indicates whether the star lies on the close or distant half of the cluster. The integral of Eq. 6 is calculated up to the $s^*$ value in order to obtain the colour excess for each star. The projection of the cluster in the sky plane is divided into square bins (see Figure 1 for a sketch of the geometry) of size $0.1 \ r_0 \times 0.1 \ r_0$. For each bin, the colour excess of all the stars in the column is averaged. The mean value is chosen to represent the colour excess in the bin.

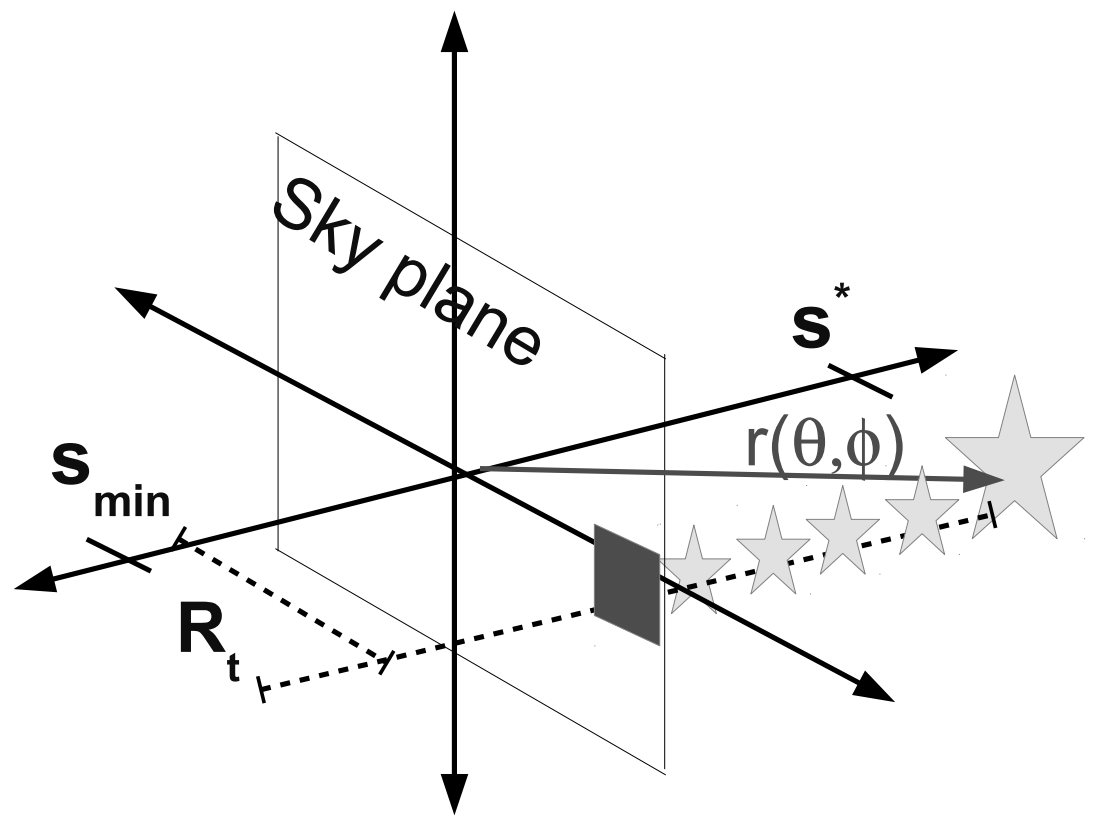

**Figure 1.** Some relevant geometrical parameters. Each star is labeled with its distance to the centre of the cluster $R_T$ and its position $s^*$ that indicates if the star is located in the close or distant half of the cluster. All stars contained in a $0.1\ r_0 \times 0.1\ r_0$ bin in the sky plane contribute to the colour excess in that bin.

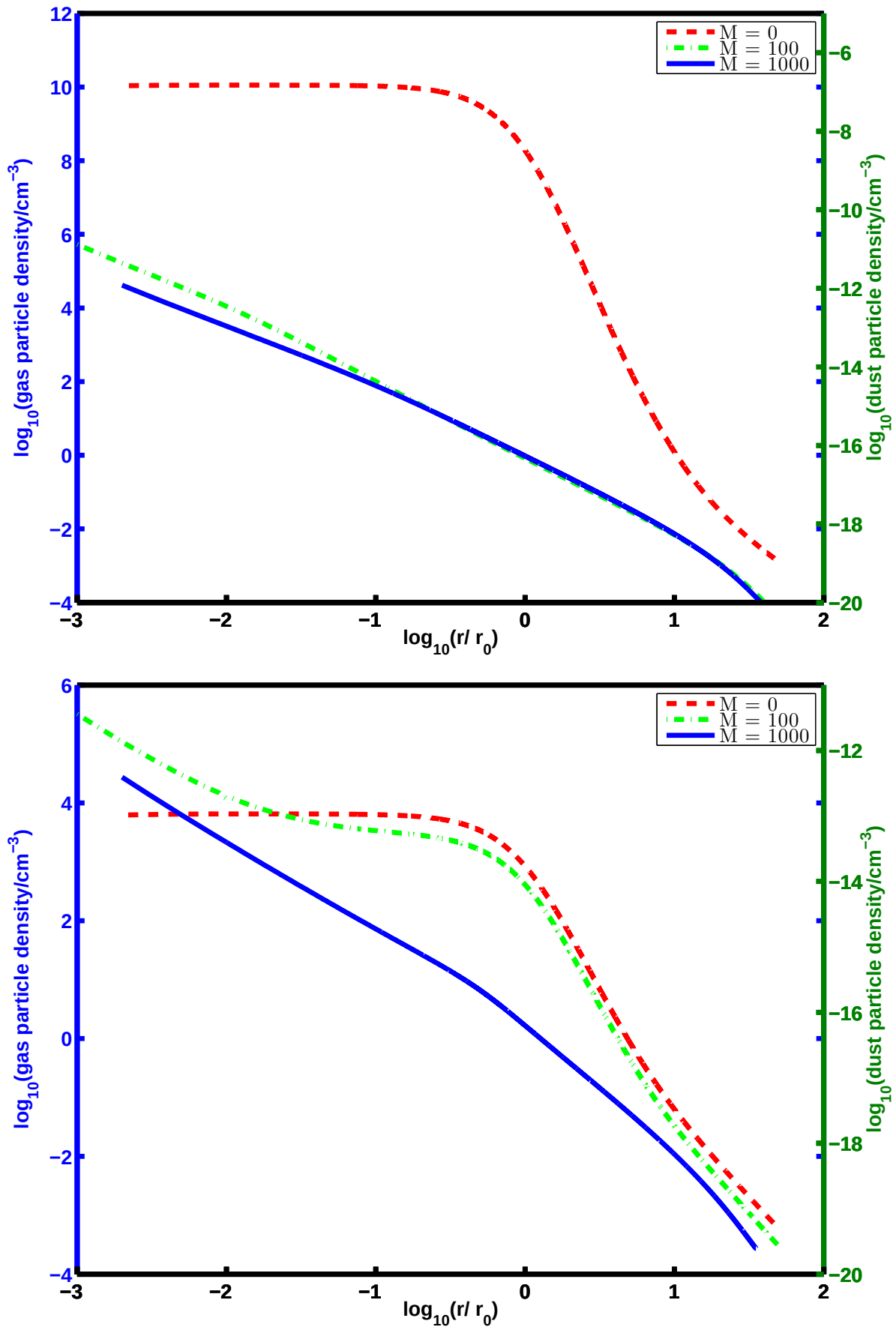

**Figure 2.** Gas and dust particle densities for M 62 from the model presented in Paper I. Top: the gas temperature is $T = 5000$K. The density obtained from the no-IMBH scenario is a few orders of magnitude higher for $r \lesssim r_0$. Bottom: the gas temperature is $T = 10000$K. The value of the density in the central areas in the no-IMBH scenario is comparable to that obtained when including the IMBH.

## 3 RESULTS

### 3.1 Gas and dust density profiles

Computing the reddening requires a known dust density profile. In order to derive it, we assume that dust grains follow the gas distribution with a density $\psi\rho_{\rm gas}$. The dust-to-gas ratio $\psi$ is metallicity-dependent and it is taken from Eq. 3 in McDonald et al. (2011). The model for the gas density was set forth in Paper I. There it was shown that, when assuming an IMBH at the centre of the cluster, two different accretion regimes (high-accretion rate regime – HAR, and low-accretion rate regime – LAR) can be found according to the value of the parameter $c_s^2/\sigma^2$, where $c_s^2$ is the sound speed in the medium and $\sigma$ is the velocity dispersion of the cluster. Here, we briefly recall the assumptions and summarize the main results of this model as follows:

(i) The gravitational pull of the stars in the cluster is considered as well as the black hole at the centre.
(ii) Winds from red giant stars act as a source of ISM.
(iii) ISM can be described as an isothermal, stationary, perfect fluid with spherical symmetry.
(iv) As a result, we find that a stagnation radius separates the inner-accretion region from the outer region, where the ISM escapes as a wind.
(v) Also, a clear correlation between the location of the stagnation radius and the black hole mass is found for those clusters in the LAR regime.

In Paper I we studied the dependence of the accretion rate on the relevant parameters, the black hole mass $M_{\rm BH}$ and the gas temperature $T$. Although several combinations of these two parameters were explored, in this work we only show the results for $T = 5000$ and $10000$ K[1], in combination with three different black hole masses ($M_{\rm BH} = 0, 100$ and $1000\ {\rm M}_\odot$) as representative values of the possible masses of IMBHs according to observations. In this work, we present the resulting gas density profiles for the first time. In addition, dust particle density profiles are calculated assuming a dust particle radius of $a = 0.01\ \mu$m (see Discussion below).

Note that the equilibrium temperature of dust particles exposed to the radiation of three blackbodies of $T_{\rm star} = 3000$, 4000 and 7500 K [2], is $\sim 12 - 20$ K, depending on grain size (Draine & Lee 1984). Further, given the gas temperature $T$ and the grain size $a$, the time it takes the dust to couple to the gas can be calculated as $t \sim m_{\rm dust}/K_{\rm s}$, where

$$K_{\rm s} = \frac{4}{3}\pi\rho_{\rm gas}a^2\sqrt{\frac{8k_{\rm B}T}{\pi\mu}}, \qquad (7)$$

is the drag coefficient, $k_{\rm B}$ is the Boltzmann constant and $\mu$ is the mean-molecular weight (Booth, Sijacki & Clarke 2015). This stopping time needs to be smaller than the crossing time of the fluid in the GC in order to obtain a dynamical coupling between dust and gas. This condition is satisfied inside the core radius, where most of the extinction takes place, for all models. Further, Lakićević et al. (2012) have shown that the detachment occurs when $\rho_{\rm gas}$ falls below $10^5$ cm$^{-3}$. This is the case for all the models for which actually compute the reddening in Section 3.3.

[1] Intermediate and higher values of the temperature were also explored but we show only two temperatures for simplicity in Figure 2.

[2] These temperature values are in range with the surface temperature of main sequence and red giant stars in GCs.

Figure 2 shows the gas and dust particle density for the particular case of M 62. The fractional mass injection rate[3] of the red giants has been set to $10^{-11}$ yr$^{-1}$ (see Paper I for a discussion on the impact of such a choice). We find that the density profile changes qualitatively from flat to steep when we consider the IMBH. Additionally, we find that for those GCs with the highest velocity dispersion parameter, such as M 62 (along with 47 Tuc, $\omega$ Cen, M 15, NGC 6440, NGC 2808, M 54, NGC 6388 and NGC 6441), if the gas temperature $T$ is low enough, the central values of the density are 2–3 orders of magnitude higher than in the IMBH scenario and this has a remarkable impact, as we shall show below.

### 3.2 Mass estimations

From the gas densities predicted by the model in Paper I, we can estimate the gas mass content in each cluster. This allows us to place constraints on the parameter space to explore, for a given cluster. In Fig. 3, we show the gas mass fraction (relative to the total cluster stellar mass) in the cluster core as a function of both, the black hole mass $M_{\rm BH}$ and $T$. Given that stars are injecting gas at a fractional rate $\alpha = 10^{-11}$ yr$^{-1}$ and considering $10^8$ yr for the time in between passages through the Galactic plane (Odenkirchen et al. 1997), it implies $10^2$ $M_\odot$ of gas for a typical cluster of $10^5$ $M_\odot$. This value is in agreement with the predictions of stellar evolution theories (see e.g. Tayler & Wood 1975). We assume this fractional content of gas $M_{\rm gas}/M_{GC} = 10^{-3}$ as an upper limit to this parameter.

It can be seen that for 47 Tuc, $\omega$ Cen, M 15, NGC 6440, NGC 2808, M 54, M 62, NGC 6388 and NGC 6441, certain combination of parameters lead to unrealistic values of the mass since they require comparable (or even higher) masses to the cluster mass itself and/or it takes the system more than $10^8$ yr to inject that amount of gas into the medium. Recall that our modelling of the gas is stationary and, as such, it requires an infinite amount of mass which is not physically acceptable. Hence, this model is only reliable for a certain combination of parameters, which we constrain here. Those combinations of parameters that lead to unrealistic physical predictions are disregarded in the following analysis. Additionally, there are observational estimations of the gas content in some of the clusters in 3. These are shown as a dashed-dotted line, when available.

As a result of our gas modelling we can place even stronger constraints on the parameters. We find a critical temperature $T_{\rm crit}$ for all clusters ranging from $\sim$ 6000 K to $\sim$ 11000 K above which there is no discrepancy between predictions and observations. The existence of $T_{\rm crit}$ can be explained by one of the following scenarios: either the ISM thermalized at temperatures higher than $T_{\rm crit}$ or, if it is cooler, only scenarios including an IMBH are compatible with the current estimations of the gas mass content in globular clusters. The extreme is NGC 6441, where even higher temperatures ($T_{\rm crit} \sim 12500$ K) are needed to reconcile the model prediction with the observational estimations (see Figure 3). In this case, the IMBH hypothesis becomes the most appealing scenario since such high temperatures are unlikely to prevail in these stellar systems (see Section Discussion below). Therefore, a precise estimation of the gas temperature in globular clusters would be very useful in determining the presence of IMBHs in GCs. We focus on the small subset of models in agreement with observations in the following.

### 3.3 Extinction maps

Extinction maps for the GCs listed in Table 1 were constructed following the procedure described in the previous section. This selection corresponds to the list of GCs for which searches of IMBHs have been reported. In Figure 4 and 5 we show the constructed extinction and dispersion maps, respectively, for M 62. The blank bins indicate the absence of stars in that bin and the colour code corresponds to log(E(B − V)). The top panels correspond to a gas temperature of $T$ = 8000 K while $T$ = 12500 K is assumed in the bottom figures. The corresponding IMBH masses are $M_{\rm BH} = 0, 100$ and 1000 $M_\odot$, from left to right. These are representative values of the mass range emerging from observations[4]. Although we show a small subset of the $M_{\rm BH} - T$ parameter space, intermediate and higher values were also explored.

We find that, if the temperature is close to the critical value $T_{\rm crit} \sim 8000$ K obtained from gas mass estimations, our prediction of the extinction in the no-IMBH scenario is incompatible with the observations of differential reddening in this cluster. Alonso García et al. (2011) constructed a 2-D extinction map of M 62 and they found a 0.2 mag deviation ($\log(\Delta{\rm E(B-V)}) \sim -0.7$) from their zero-point reddening value, E(B-V) = 0.42 mag. Considering that this estimation also includes foreground reddening, it is not straightforward to compare our maps with theirs. Further, the authors state that the increase in the extincion in the east-west direction south from the cluster centre is due to material (cloud) lying along the line of sight but outside the cluster. However, note that their zero-point reddeing value is located almost coincident with the cluster centre and that there seems to exist some radial symmetry (decreasing outwards), if the cloud is neglected. Our results for the model with $T_{\rm crit} \sim 8000$ K and no black hole, reproduces this feature but it does not reproduce the 0.2 mag deviation observed by Alonso García et al. (2011), since we estimate a maximum deviation of 0.31 mag from the centre outwards. Nonetheless, adding the black hole decreases both the gas and dust content in the cluster, leading to smaller values of the intrinsic extinction. Given that these values (see the rest of the maps in Fig. 4) are smaller than the resolution in the map of Alonso García et al. (2012) and that foreground extincion could be variable, we can not rule out these models. Hence, if gas temperature is close to 8000 K only models including an IMBH are not in contradiction with the observations, favouring the IMBH hypothesis. As gas temperature increases, the gas and dust content decrease and so does the extinction in M 62. We find that a temperature only 1000 K higher, reduces the core density enough to avoid discrepancies between our predictions and observations. The lower pannel of Fig. 4 shows the maps for $T = 10000$ K for reference. For these hotter temperatures, both scenarios, with and without an IMBH seem plausible.

[3] The mass injection rate is defined such that $dM_{cluster}/dt = \alpha M_{cluster}$.

[4] Although the observations in $\omega$ Cen have led to much higher masses, those high values can only be considered for such a massive GC as this one.

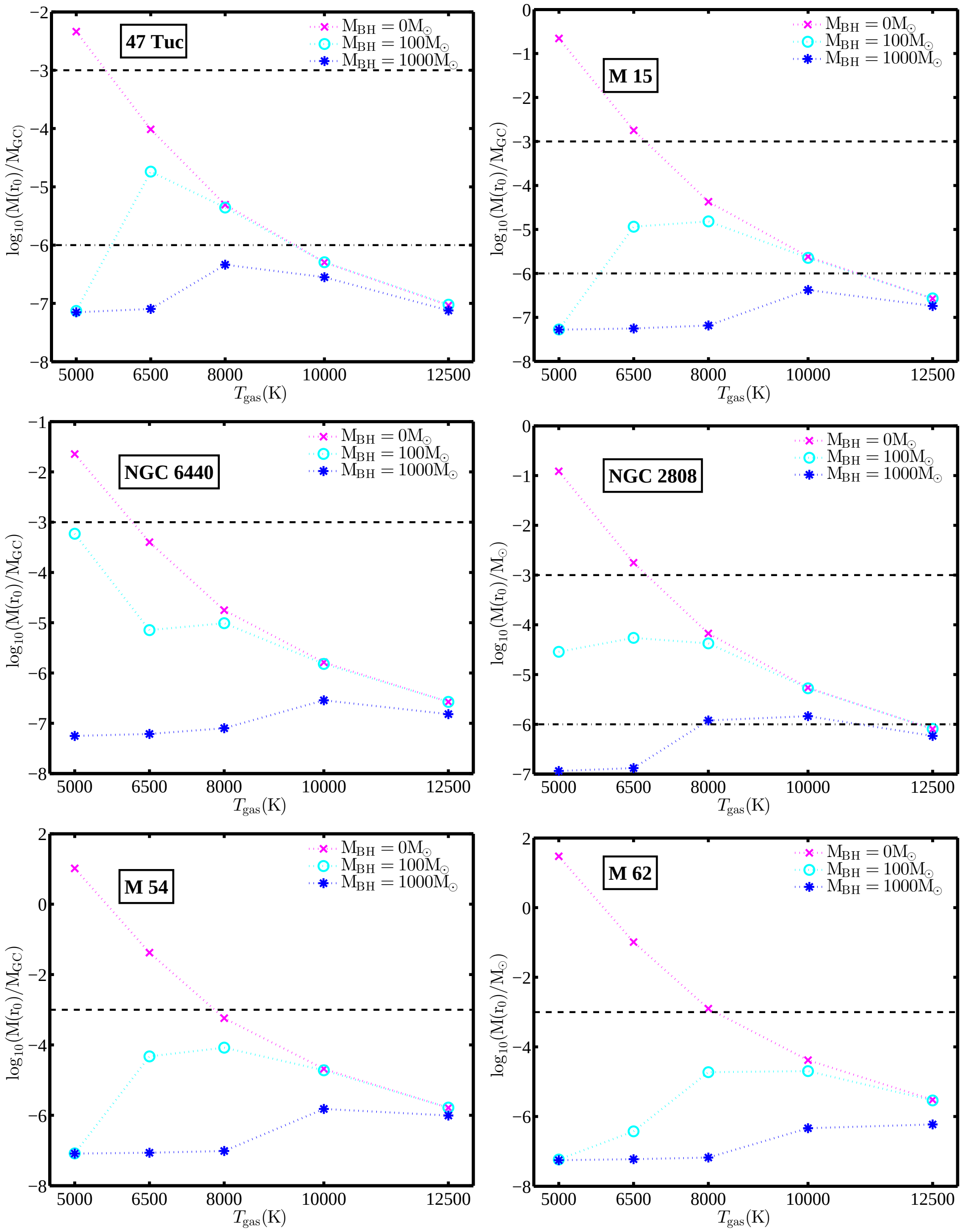

**Figure 3.** Gas mass–fraction inside the core radius for a subset of our sample of GCs. For low temperatures, if the black hole is not present the predicted values are above the $10^{-3}$ limit (dashed line) derived from stellar evolution theories. Dashed-dotted line shows the observational limit, when available. The corresponding references are: Freire et al. (2001) for 47 Tuc, van Loon et al. (2006) for M 15, Lynch et al. (1989) for NGC 6388, Bowers et al. (1979) for NGC 6441 and McDonald et al. (2009) for $\omega$ Cen.

The same kind of maps were constructed for $\omega$ Cen (see Fig. 6). For this cluster, the smallest dust particles are considered, with $\sim 3.5$Å radius. According to the results shown in Fig. 3, the critical temperature is $T_{\rm crit} \sim 9000$ K for *all* masses, placing strong constraints on the gas temperature in this cluster. Additionally, we find that if $T \geqslant T_{\rm crit}$ there is no efficient accretion onto the black hole and models with and without the black hole lead to similar gas density profiles and mass content. Therefore, the dust distribution and, consequently, the extinction, is the same for all masses. van Loon et al. (2009) constructed 2-D maps of the extinction in $\omega$ Cen. They found a cloud in the periphery of the cluster and a curious dust cloud near the centre. This excess of reddening towards the centre (of $\sim$ 0.024 mag, $\log(\mathrm{E(B-V)}) \sim -1.6$) due to a foreground cloud dominates the colour excess and it hinders the analysis of the intracluster extinction. Further, our predictions are below the sensitivity of the observed map and the model can not be discarded nor supported.

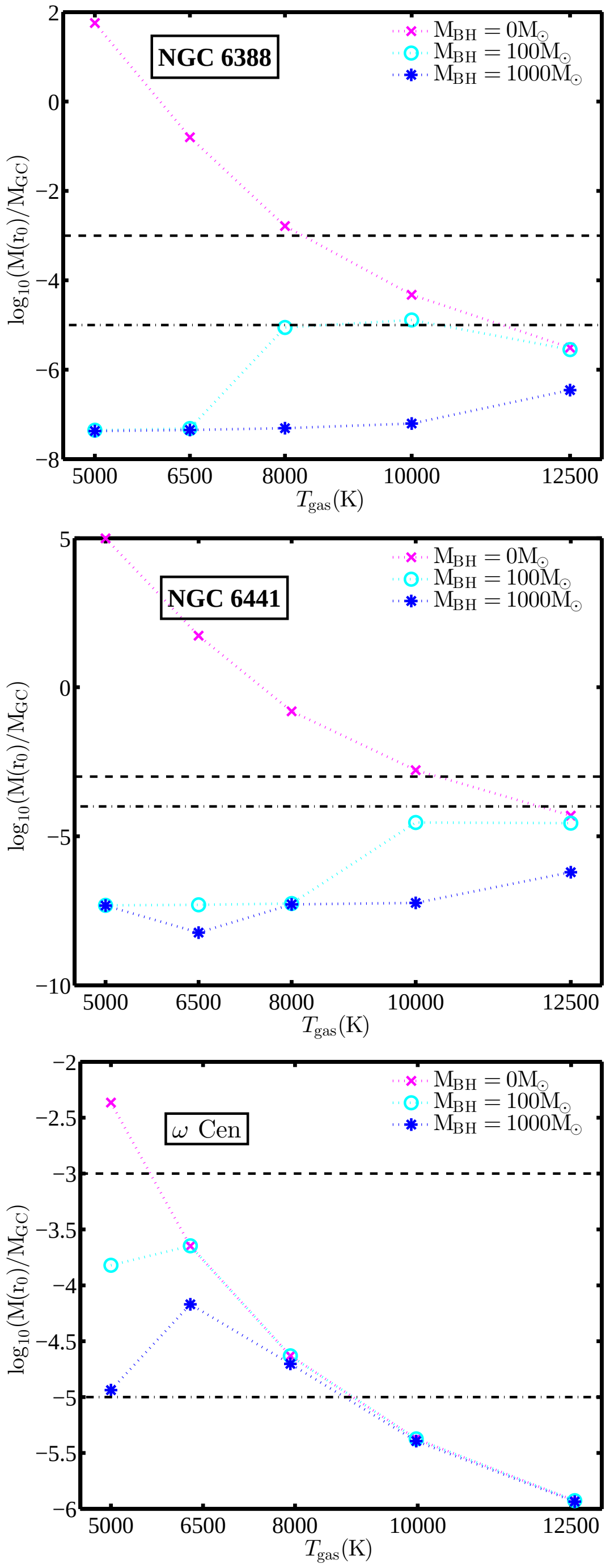

**Figure 3.** Cont.

For the rest of the clusters, we only have available the reddening towards the cluster centre, obtained from integrated colours and colour-magnitude diagrams where the light is dominated by the central stars (Harris 1996). However, this is not useful to our aims, since we are interested in studying the extinction inside the cluster and we need spatial variations of the reddening to make predictions with our modelling. We did check for all clusters that the mean value of the extinction of the stars inside the core radius is below[5] the estimations of Harris (1996). Unfortunately, no further analysis is possible in these cases.

## 4 DISCUSSION

The question "What happened to intracluster medium?" still does not have a satisfactory answer. While different scenarios have been proposed by several authors, still none of them prevail over the others. The only consensus is that the intracluster medium is being removed from the cluster. In this work we explore the possibility that the ISM is being accreted by an IMBH. Although its presence at the centre of GCs is still under debate, observational works have led to place constraints on the mass of the putative black hole in certain GCs, by means of stellar dynamics or accretion detection. However, little has been done investigating their effects on the surrounding media. This was the main purpose of the model developed in Paper I. In this work we combine these two issues. On one hand, a new mechanism for the removal of the ISM is proposed. On the other, the results obtained allow us to decide whether the model containing the IMBH or not suits better the observations. This is our main goal: to make predictions about the presence of an IMBH. If the results suggest the need to invoke an IMBH, this strengthens the hypothesis that accretion onto an IMBH is a possible and valid cleansing mechanism in GCs.

Our main conclusion is that, for some GCs (namely, M 62, NGC 2808, M 15, 47 Tuc, $\omega$ Cen, M 54, NGC 6440, NGC 6441 and NGC 6388), if the gas temperature is below a critical value $T_{\rm crit}$, the gas mass content is orders of magnitude above the observational estimation if the IMBH is not present at the centre of the cluster. However, for the same temperature, adding an IMBH reduces the density value and the total gas mass can be reconciled with the observations. This feature can be well understood in terms of the gas energetics: for lower temperatures, in abscense of an accreting IMBH, the gas lacks the required energy to escape as a wind and is retained in the cluster potential. As the temperature goes up, the gas avoids its retention in the cluster potential and, hence, the density decreases and this explains the differences in the total gas mass in the cluster. As discussed in the previous section, this subset of clusters corresponds to those with velocity dispersion $\sigma > 10$ km s$^{-1}$. Clearly, the detection of the ISM and the determination of its temperature would be crucial to confirm the existence of an IMBH in these GCs and, therefore, to support the hypothesis of accretion as an efficient cleansing mechanism. However, as pointed out by different authors (e.g., Scott & Rose 1975, and references therein), the gas temperature should depend only on the properties of the radiation field of the cluster. The ISM of GCs with an important source of UV radiation is expected to be ionized at $T \gtrsim 10^4$K. From the GCs listed above, only four (47 Tuc, NGC 2808, M 15 and M 62) have stars hot enough to provide a wealth of UV radiation (log T > 4.3) (O'Connell et al. 1997; Miocchi 2007)[6]. The rest of them (M 54, NGC 6388, NGC 6440 and NGC 6441) lack these stars and, hence, become the best candidates to search for an IMBH. NGC 6441 is the most appealing one as our models predict a high critical temperature, near 12500 K.

[5] Actually, our estimations are at least 2 orders of magnitude below that limit.

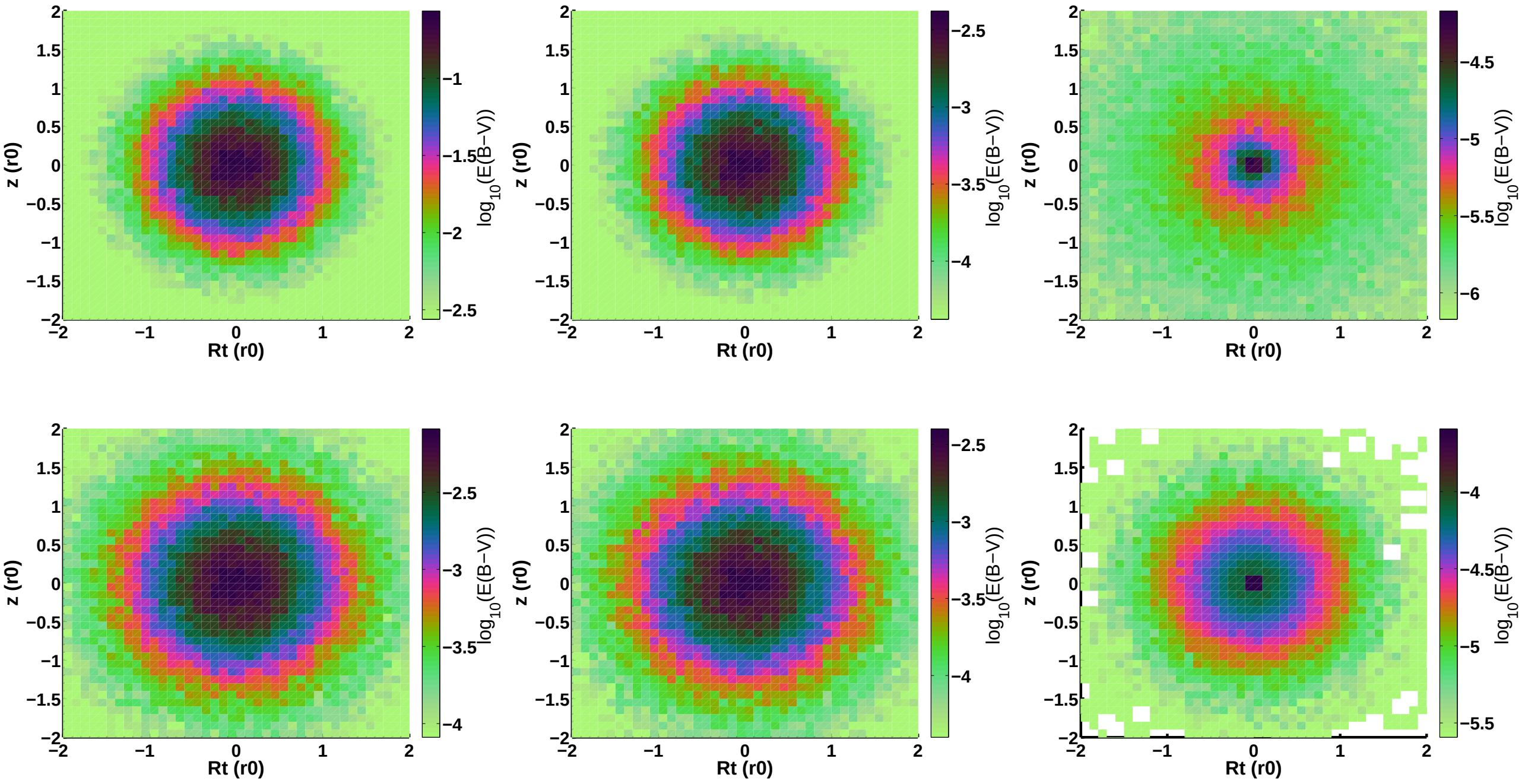

**Figure 4.** Extinction maps for M 62 from the model presented in Paper I. In both upper and lower panels the corresponding masses are $M_{\rm BH} = 0, 100$ and $1000\ M_{\odot}$, from left to right. Top: the gas temperature is $T \sim 8000$K. Bottom: the gas temperature is $T = 10000$K. Note the different colour scaling through the plots.

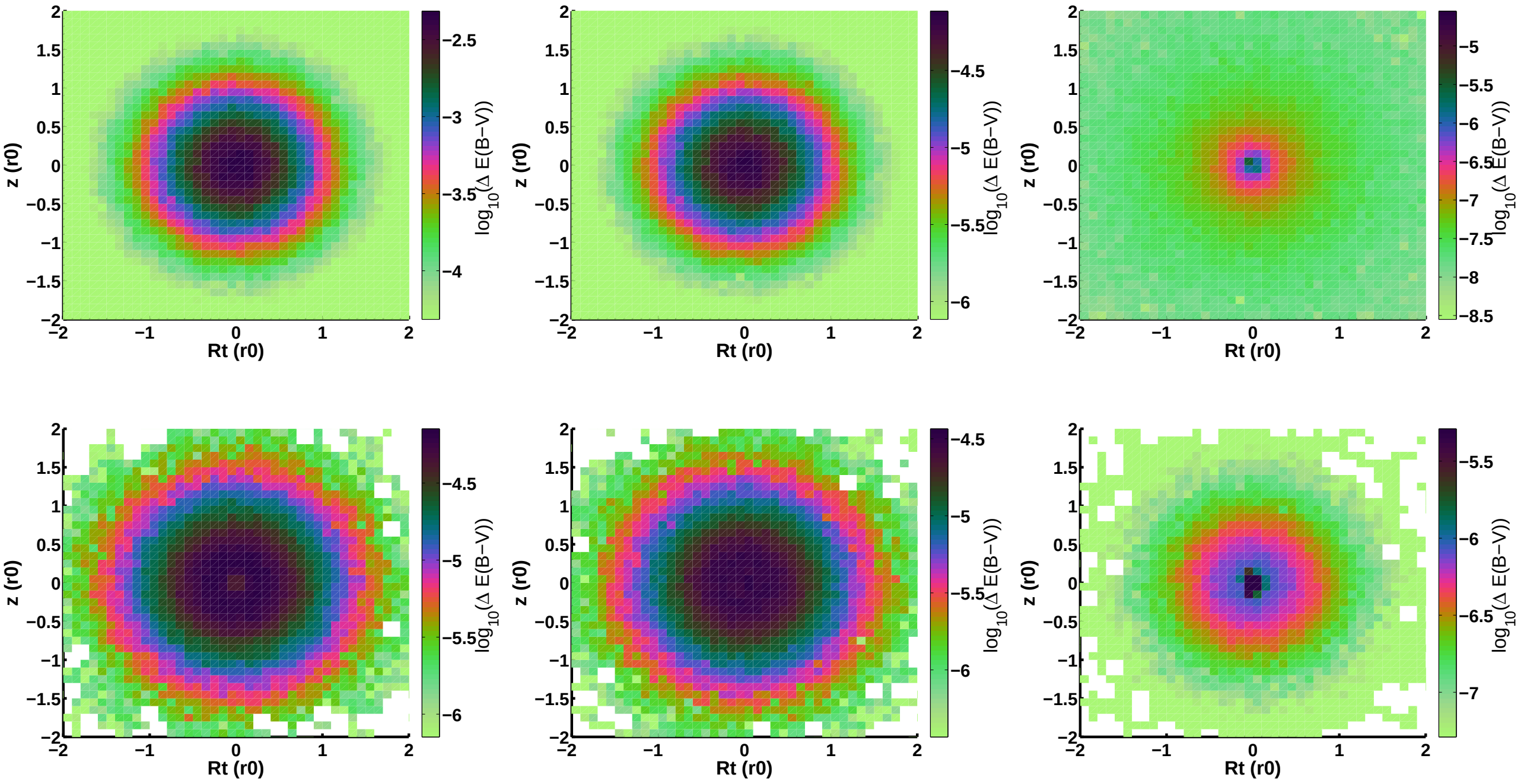

**Figure 5.** Dispersion of the extinction maps for M 62 from the model presented in Paper I. In both upper and lower panels the corresponding masses are $M_{\rm BH} = 0, 100$ and $1000\ M_{\odot}$, from left to right. Top: the gas temperature is $T \sim 8000$K. Bottom: the gas temperature is $T = 10000$K. Note the different colour scaling through the plots.

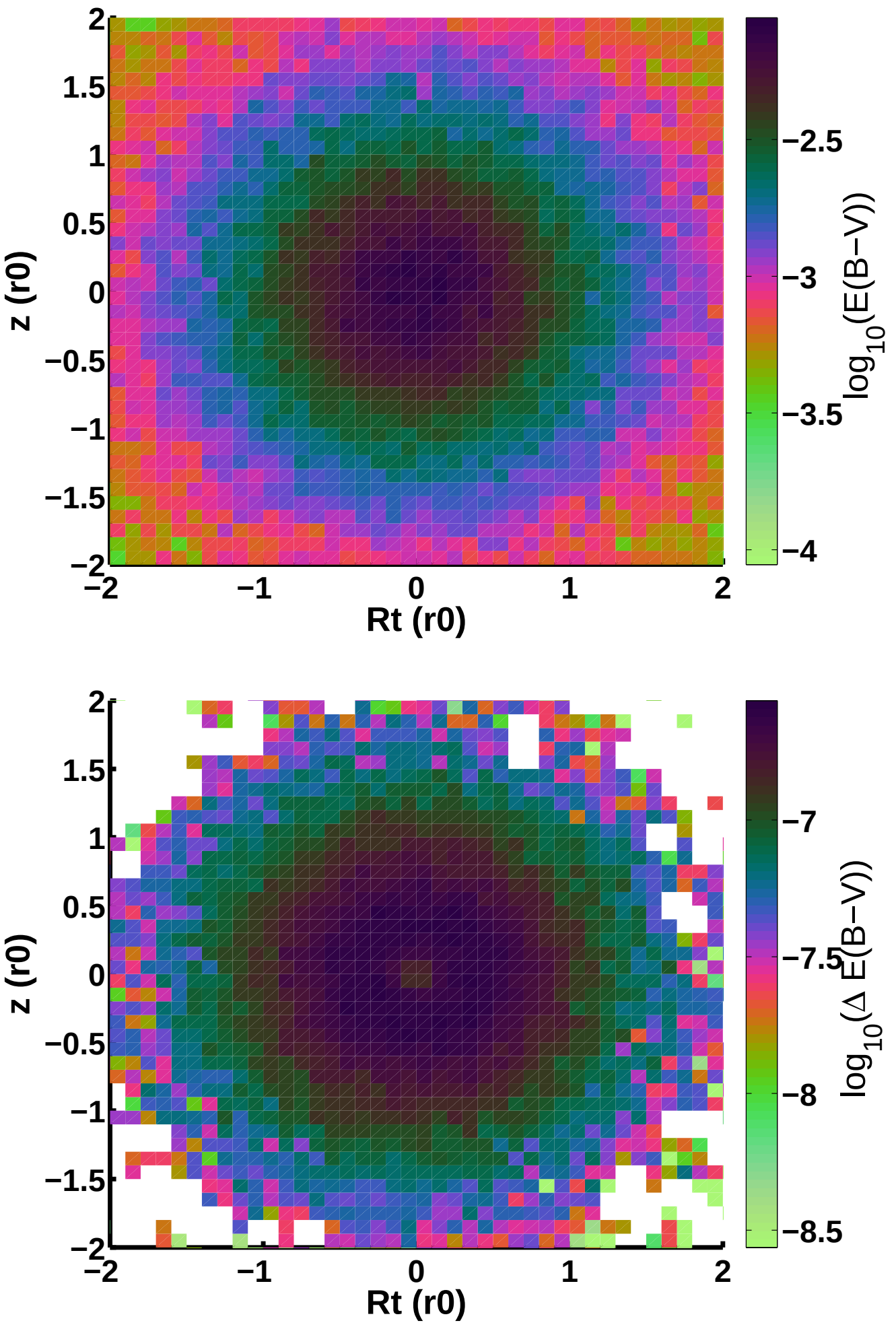

**Figure 6.** Extinction and dispersion map for $\omega$ Cen from the model presented in Paper I. It corresponds to a model with no IMBH and T = 10000 K. For temperatures $T \gtrsim 9000\ K$ adding the black hole does not change the mass distribution.

[6] For a detailed list of sources in 47 Tuc and an estimation of the flux needed to ionize its medium, see McDonald & Zijlstra (2015). McDonald et al. (2015) present another approach about the effect of UV radiation on the 47 Tuc medium.

In Paper I, we have discussed the impact of the fractional mass injection rate, $\alpha$ on our results. Since the gas mass depends linearly on this parameter, a brief discussion on this dependence seems relevant. The value of $\alpha$ has been constrained to the range $10^{-14} - 10^{-11}$ yr$^{-1}$ from observational works (Scott & Rose 1975; Fusi Pecci & Renzini 1975; Dupree et al. 1994; Mauas et al. 2006; Priestley et al. 2011). In Paper I, we managed to estimate $\alpha$ by means of comparing our model with the traditional Bondi & Hoyle (1944) model and using the current observational constraint on the ISM obtained by Freire et al. (2001) in 47 Tuc. This yielded $\alpha$ values in agreement with the range stated before. Nonetheless, even if $\alpha$ was lower, it would not suffice to reconcile the estimated gas mass with the observational constraints since, in most cases, the difference is greater than the 3 orders of magnitude of span in $\alpha$.

We also constructed synthetic exctinction maps assuming the dusty component of the ISM couples to the gaseous component, with a density $\psi$ times lower, depending on the cluster metallicity. It is worth pointing out that any spherical density distribution can be tested by means of this method. The maps were constructed on a star-by-star basis and consider only the extinction produced by the dust *within* the cluster. van Loon et al. (2009) and Alonso García et al. (2012) constructed 2-D extinction maps for $\omega$ Cen and M 62, respectively. In the case of $\omega$ Cen, we can draw no conclusions about the accuracy of our extinction modelling in this cluster, given the current observational precision in extinction maps. On the other hand, our map for M 62 with $T = 8000$ K and no black hole, reproduces the spatial distribution observed by Alonso García et al. (2012), although it overpredicts the difference in magnitudes between the central area of the cluster and the periphery. This result suggests that a model with an IMBH suits better the observations of the reddening in this cluster, if gas temperature is close to 8000 K. Of course, this result depends on the dust particle size; note that Eq. 6 scales as $\sim a^{-1}$. Recall that for M 62 we adopted a value of $a \sim 0.01\mu$m. In Fig. 7 we show 1D cuts of the upper-panel maps in Fig. 4 as dashed-coloured lines. The shaded regions account for the spread in the grain size. It can be seen that for smallest particles, the extinction can be enhanced a few orders of magnitude. Constraining the dust grain sizes as well as the gas temperature would be of great relevance to the study of IMBHs via dust extinction mapping.

## 5 CONCLUSIONS

In this work we have developed a method that allow us to decide whether the presence of an IMBH at the centre of a GC is compatible with current estimations of the colour excess, $E(B-V)$. As stated before, the construction method of the maps is independent of the density profile used. Hence, this procedure can be applied to any density distribution of interest. Unfortunately, observational techniques still include the reddening produced along the path between the observer and the cluster while our method only estimates the reddening due to intracluster dust. This avoids a detailed comparison between our maps and the observational maps of either van Loon et al. (2009) or Alonso García et al. (2011). This comparison would be useful, for example, to constrain the mass of the IMBH. However, we can explain the spatial symmetry observed by these authors in M 62 and $\omega$ Cen, under certain natural assumptions about the model parameters. Furthermore, in M 62, if gas temperature is close to 8000 K, our results suggest that an IMBH needs to be invoked, given the current observations of the extinction in this cluster. Finally, the cluster sample of Alonso García et al. (2012) corresponds to GCs with low Galactic latitude, where foreground extinction is highly variable and might obliterate the effect due to intracluster dust (M. Catelán, private communication). To overcome this issue, high-latitude clusters extinction maps should be measured and, if that is the case, the study of the extinction in GCs could be an extra observational method to search for IMBHs that complements the current ones based on star dynamics and accretion emission.

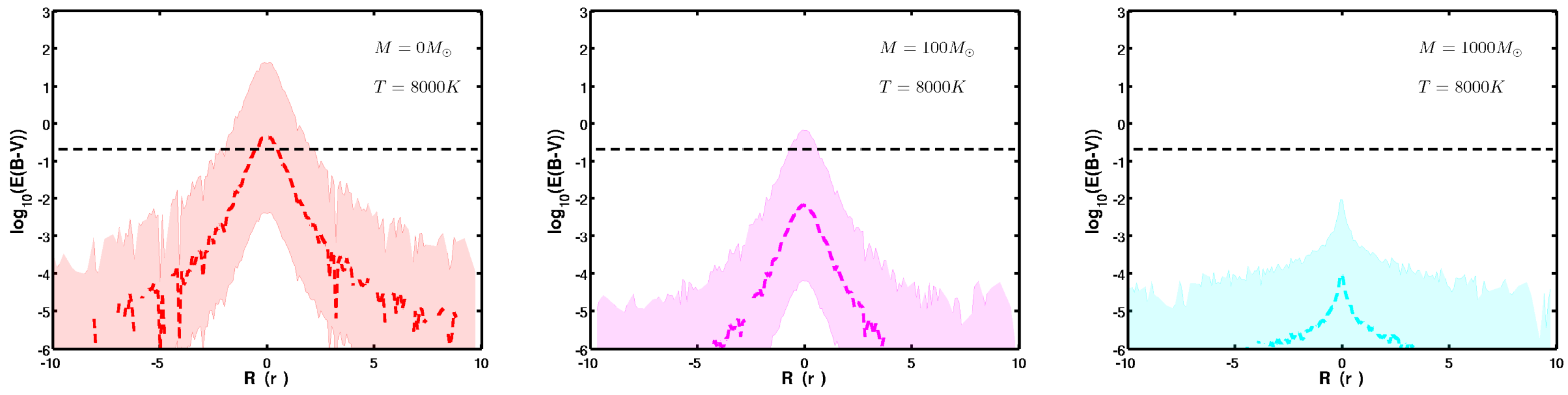

**Figure 7.** 1-dimensional cuts of the extinction maps for M 62 from the model presented in Paper I (colour dashed-lines). The gas temperature is $T \sim 8000$K and the corresponding masses are $M_{\rm BH} = 0, 100$ and 1000 $M_\odot$, from left to right. The shaded areas cover the uncertainty in the extinction value due to the dispersion in the dust grain size. The horizontal dashed black line shows the $\sim$ 0.2 mag difference in Alonso García et al. (2011) map for this cluster.

## ACKNOWLEDGEMENTS

C.P. wants to express her gratitude to the referee, Jacco van Loon, for such a detailed evaluation of the manuscript. All the suggestions only improved our work.